\documentclass[letterpaper]{article} 
\usepackage[draft]{aaai25}  
\usepackage{times}  
\usepackage{helvet}  
\usepackage{courier}  
\usepackage[hyphens]{url}  
\usepackage{graphicx} 
\def\UrlFont{\rm}  
\usepackage{natbib}  
\usepackage{caption} 
\usepackage{algorithm}
\usepackage{algorithmic}
\usepackage{float}

\usepackage{newfloat}
\usepackage{listings}
\DeclareCaptionStyle{ruled}{labelfont=normalfont,labelsep=colon,strut=off} 
\floatstyle{ruled}
\newfloat{listing}{tb}{lst}{}
\floatname{listing}{Listing}
\usepackage{mathtools}

\title{Physically Based Neural Bidirectional Reflectance Distribution Function}
\author{
    Chenliang Zhou, Alejandro Sztrajman, Gilles Rainer, Fangcheng Zhong\\Fazilet Gokbudak, Zhilin Guo, Weihao Xia, Rafal Mantiuk, Cengiz Oztireli
}
\affiliations{
    Department of Computer Science and Technology\\
    University of Cambridge\\
    chenliang.zhou@cst.cam.ac.uk 
}

\usepackage{xcolor}

\usepackage{amsmath}
\newcommand{\intd}{\mathrm{d}}
\renewcommand{\vec}[1]{{\boldsymbol{\mathbf{#1}}}}

\RequirePackage{xspace}
\makeatletter
\DeclareRobustCommand\onedot{\futurelet\@let@token\@onedot}
\def\@onedot{\ifx\@let@token.\else.\null\fi\xspace}

\def\eg{\emph{e.g}\onedot} 
\def\ie{\emph{i.e}\onedot}

\makeatother

\usepackage[capitalize]{cleveref}
\crefname{section}{Sec.}{Secs.}
\Crefname{section}{Section}{Sections}
\Crefname{table}{Table}{Tables}
\crefname{table}{Tab.}{Tabs.}

\usepackage{subcaption}
\usepackage{amsfonts}
\usepackage{booktabs}

\begin{document}

\maketitle

\begin{abstract}
We introduce the \emph{physically based neural bidirectional reflectance distribution function~(PBNBRDF)}, a novel, continuous representation for material appearance based on neural fields. Our model accurately reconstructs real-world materials while uniquely enforcing physical properties for realistic BRDFs, specifically Helmholtz reciprocity via reparametrization and energy passivity via efficient analytical integration. We conduct a systematic analysis demonstrating the benefits of adhering to these physical laws on the visual quality of reconstructed materials. Additionally, we enhance the color accuracy of neural BRDFs by introducing \emph{chromaticity enforcement} supervising the norms of RGB channels. Through both qualitative and quantitative experiments on multiple databases of measured real-world BRDFs, we show that adhering to these physical constraints enables neural fields to more faithfully and stably represent the original data and achieve higher rendering quality.
\end{abstract}




\section{Introduction}
\label{sec:intro}
Representational learning has become the standard method for modeling complex spatial distributions and functions from measured data in computer graphics and vision. \emph{Implicit neural representations (INRs)} utilize neural fields, such as multi-layer perceptrons~(MLPs), to estimate functions that represent a signal continuously by training on discretely sampled data. These representations have applications ranging from 2D images~\cite{sitzmann2019siren} to directional reflectance functions~\cite{Rainer2019Neural, sztrajman2021nbrdf}, 3D surfaces~\cite{wang2021neus}, and scene density and radiance~\cite{mildenhall2020nerf, RBRD22}. They have demonstrated superior expressiveness, interpolation ability, and fidelity to the underlying data compared to analytic parametric models. However, this expressiveness comes with a lack of constraints on the function learned by the network. This issue is often addressed through the use of priors~\cite[\eg,][]{perez2023building}, minimizing total variation~\cite{yeh2022total}, or incorporating inductive biases relevant to the problem~\cite[\eg,][]{wang2021neus}.

In the domain of material appearance modeling, the most common function used for rendering is the \emph{bidirectional reflectance distribution function~(BRDF)}~\cite{Nicodemus1977GeometricalCA}. Implicit neural representations for BRDF models (\eg, \emph{neural BRDFs~(NBRDFs)}~\cite{sztrajman2021nbrdf}) have proven efficient in modeling material appearance due to their expressiveness and compactness, coupled with high fidelity to real-world training data. However, BRDFs are derived from first principles and must obey strict physical constraints: Helmholtz reciprocity ensures that reflected and absorbed energy are independent of the direction of light travel, and energy passivity ensures that a point cannot reflect more light than it receives (see \cref{sec:background} for more details). Parametric BRDF models~\cite{brdfsurvey} inherently respect these constraints, whereas NBRDFs, optimized to match measured data accurately, are not guaranteed to comply. This non-compliance may lead to significant artifacts in physical simulations of light transport, such as path tracing, where BRDFs are expected to satisfy strict physical constraints (\eg, \cref{fig:nbrdf-bad,fig:energy_renderings}).

In this paper, we propose the \emph{physically based neural BRDF~(PBNBRDF)}, which combines multiple novel methods to ensure physical plausibility and improve the perceptual fidelity of the underlying BRDFs. Through qualitative and quantitative evaluations, we demonstrate the effectiveness of our methods for creating physically sound and accurate neural BRDFs. We highlight the necessity of physically based neural fields to bridge the gap between well-established models like BRDFs and real-world data.

\section{Related Work}
\label{sec:relwork}
\paragraph{Material Acquisition and Databases} Traditionally, material reflectance has been measured using gonioreflectometers, devices that control the light and view direction to measure reflected radiance~\cite{White98, gonio}. Standard setups include arrays of cameras~\cite{weinmann2015advances} and light sources~\cite{haindl2013visual}. Other configurations also exist, ranging from specific geometries for measuring various values~\cite{marschner99, kaleidoscop} to setups capable of capturing entire material patches~\cite{danabtf, weinmann-2014}. These scanners have enabled the creation of well-known BRDF databases~\cite{Matusik2003datadriven, nielsen2015,rgl2018}, which we utilize in this work (see \cref{sec:results}).

\paragraph{BRDF Modeling} Replicating the appearance of real-world materials represented via bidirectional reflectance distribution functions~(BRDFs)~\cite{Nicodemus1977GeometricalCA} has been a highly prolific research field in computer graphics. Analytic models have been the most common representations for BRDFs, characterized by their ease of editing, fast evaluation speeds, and minimal memory requirements. Classic models include Phong~\cite{phong}, Cook-Torrance~\cite{cook-torrance}, Ward~\cite{ward}, Lafortune~\cite{lafortune}, GGX~\cite{ggx}, and Disney~\cite{disney}. Despite their multiple advantages, these models rely on simplified assumptions about the distributions of reflectance, resulting in a limited ability to accurately reconstruct real-world materials~\cite{ngan2005, guarnera2016}.

Data-driven representations offer a powerful alternative by directly leveraging real-world measurement data. A common approach involves using tabular data, sampling reflectance at discrete intervals of directions and wavelengths~\cite{Matusik2003datadriven}.
Dimensionality reduction techniques~\cite{lawrence2004, nielsen2015} are used to mitigate the space issues arrised from high dimensionality.

\paragraph{Neural BRDFs} Implicit neural representation presents a modern data-driven alternative, employing neural networks to learn a dense, continuous representation of BRDF data~\cite[\eg,][]{deepbrdf, zheng, sztrajman2021nbrdf, gokbudak2023hypernetworks}. In \emph{neural BRDF~(NBRDF)}~\cite{sztrajman2021nbrdf}, a lightweight neural field is overfitted to a given material by training on the measured BRDF.
For layered BRDFs, a universal decoder is used to ensure the BRDF descriptors all exist in the same latent space~\cite{Fanlayered}. MetaLayer~\cite{metalayered}, on the other hand, learns the mapping from parameters of explicit BRDF models into this neural domain.

Methods have been proposed to improve the sampling pattern~\cite{liuLearningLearnSample2023, soh2023neural}, enforce hard constraints of arbitrary differential order~\cite{zhong2023neural}, and accelerate the fitting process~\cite{fischer2022metappearance}. Other efforts have focused on achieving real-time rendering with neural BRDFs by optimizing the parameterization~\cite{sphericalbrdf} or integrating with hardware~\cite{zeltnerneural}.

Neural BRDF models excel in accuracy and efficiency, providing a scalable solution for a broad spectrum of materials while maintaining a compact memory footprint. However, a significant limitation of neural representations is their lack of compliance with two fundamental BRDF physical constraints: Helmholtz reciprocity and energy passivity. This non-compliance can result in physically implausible outcomes and undesirable artifacts in rendered images, highlighting the need for additional mechanisms to ensure physical validity and realism.

\section{Background}
\label{sec:background}

\subsection{Bidirectional Reflectance Distribution Functions}
\label{sec:background:brdf}
When a beam of light reaches the surface of a material, its intensity is scattered according to a distribution determined by the complex interaction between the material and the light. This interaction is described by the \emph{bidirectional reflectance distribution function~(BRDF)}~\cite{nicodemus1977}. The BRDF $f_r$ quantifies the ratio between the reflected differential radiance $\intd L_o$ and the incoming differential irradiance $\intd E_i$ at the point of incidence:
\begin{equation}
    f_r(\vec{\omega}_i, \vec{\omega}_o) = f_r(\theta_i, \varphi_i, \theta_o, \varphi_o) = \frac{\intd L_o(\vec{\omega}_o)}{\intd E_i(\vec{\omega}_i)}.
    \footnote{Note that in reality, the quantities involved, such as $f_r$, $L_o$, and $E_i$, are 3-vectors representing the RGB channels or n-vectors representing spectral bands. For simplicity, we will work with one of the channels; the full vector can be obtained by repeating the operation for each channel. If aggregation is necessary (\eg, in \cref{eq:EPL}), it will be taken as the mean of the channels.}
    \label{eq:brdf}
\end{equation}
This function depends on the incident direction $\vec{\omega}_i=(\theta_i, \varphi_i)$ and the outgoing direction $\vec{\omega}_o=(\theta_o, \varphi_o)$, expressed in spherical coordinates. Here, the polar angles $\theta_i, \theta_o \in [0, 2\pi)$ and the azimuthal angles $\varphi_i, \varphi_o \in [0, \frac{\pi}{2}]$.

For physical plausibility, BRDFs need to adhere to three fundamental properties: For any pair of directions $\vec{\omega}_i, \vec{\omega}_o$,
\begin{itemize}
    \item Positivity: $f_r(\vec{\omega}_i, \vec{\omega}_o) \geq 0$;
    \item Helmholtz reciprocity~\cite{stokes1849perfect}: $f_r(\vec{\omega}_i, \vec{\omega}_o) = f_r(\vec{\omega}_o, \vec{\omega}_i)$.
    \item Energy passivity: Reflected energy cannot exceed incident energy; that is,
    \begin{equation}
        \int_\Omega f_r(\vec{\omega}_i, \vec{\omega}_o)\cos{\theta_o}\intd \vec{\omega}_o \leq 1,
    \label{eq:energy}
    \end{equation}
    where $\Omega$ is the upper hemisphere. Note that it is valid for the reflected energy to be strictly less than the incident energy (left-hand side is strictly less than 1), in which case the material absorbs the energy.
\end{itemize}

\subsection{Rusinkiewicz Parametrization}
\label{sec:Rusinkiewicz Parametrization}
An alternative parameterization of the 4D directional space $(\vec{\omega}_i, \vec{\omega}_o)$ is proposed by Rusinkiewicz~\cite{rusinkiewicz1998}, using the half and difference vectors $\vec{h}, \vec{d} \in \mathbb{R}^3$ or their spherical coordinates $\theta_{\vec{h}}, \varphi_{\vec{h}}, \theta_{\vec{d}}, \varphi_{\vec{d}}$:
\begin{align}
\begin{split}
    \label{eq:rusinkiewicz}
    \vec{h} &:= \frac{\hat{\vec{\omega}}_i + \hat{\vec{\omega}}_o}{\lVert \hat{\vec{\omega}}_i + \hat{\vec{\omega}}_o \rVert}; \\
    \vec{d} &:= R_{\vec{b},-\theta_\vec{h}} R_{\vec{n},-\varphi_\vec{h}} \hat{\vec{\omega}}_i,
\end{split}
\end{align}
where $\hat{\vec{\omega}}_i, \hat{\vec{\omega}}_o \in \mathbb{R}^3$ are the incident and outgoing directions in Euclidean coordinates, $R_{\vec{v}, \alpha}$ is the rotation around the vector $\vec{v}$ by the angle $\alpha$, $\vec{n}$ is the surface normal, and $\vec{b}$ is the surface binormal. Note that the composite rotation $R_{\vec{b},-\theta_\vec{h}} R_{\vec{n},-\varphi_\vec{h}}$ rotates $\vec{h}$ to the north pole $(0, 0, 1)$.

One advantage of Rusinkiewicz parametrization is that, under this parametrization, Helmholtz reciprocity can be simply expressed as $\pi$-periodicity in $\varphi_{\vec{d}}$:
\begin{equation}
    f_r(\theta_{\vec{h}}, \varphi_{\vec{h}}, \theta_{\vec{d}}, \varphi_{\vec{d}}) = f_r(\theta_{\vec{h}}, \varphi_{\vec{h}}, \theta_{\vec{d}}, \varphi_{\vec{d}} + \pi).
\label{eq:helmholtz}
\end{equation}


\subsection{Neural BRDFs}
\emph{Neural BRDFs~(NBRDFs)}~\cite{sztrajman2021nbrdf} are proposed as an implicit neural representation to model BRDFs from real-world measurements. An NBRDF $f_r^{\vec{\xi}}$ is a lightweight neural field parameterized by $\vec{\xi}$ and trained to overfit to a given material using the mean absolute logarithmic loss function
\begin{equation}
    \mathcal{L}_\text{NBRDF}(\vec{\xi}) := \left| \log \left(1 + f_r \cos\theta_i \right) - \log \left(1 + f_r^{\vec{\xi}} \cos\theta_i \right) \right|,
    \label{eq:nbrdf-loss}
\end{equation}
where $f_r$ is the ground-truth BRDF to which $f_r^{\vec{\xi}}$ overfits.

\begin{figure}
    \centering
    \begin{subfigure}[b]{0.35\linewidth}
        \centering
    \includegraphics[height=2.9cm]{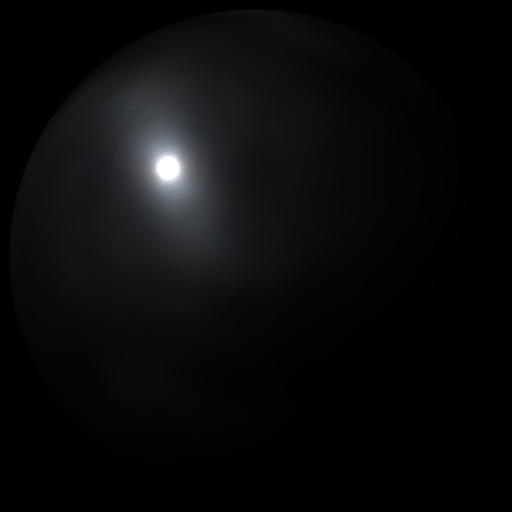}
        \caption{Ground truth}
    \end{subfigure}~
    \begin{subfigure}[b]{0.35\linewidth}
        \centering
    \includegraphics[height=2.9cm]{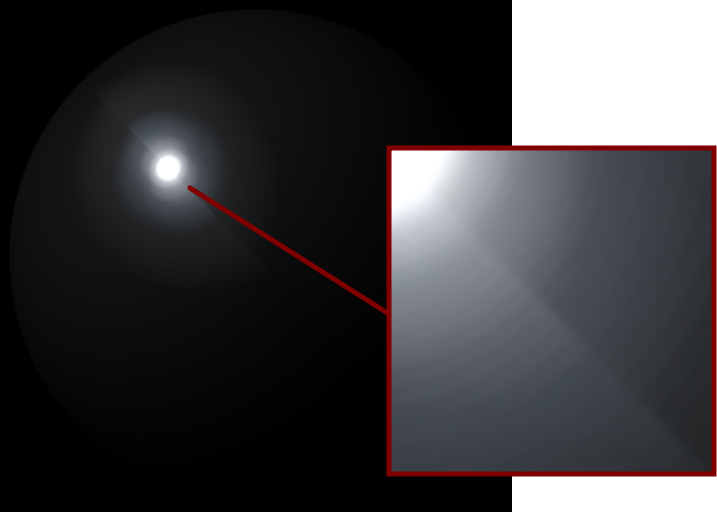}
        \caption{NBRDF}
    \end{subfigure}
    \caption{NBRDF~\cite{sztrajman2021nbrdf} violates Helmholtz reciprocity, leading to tangential discontinuities in the specular region (see diagonal discontinuity in the close-up inset of the zoomed-in image).}
    \label{fig:nbrdf-bad}
\end{figure}

However, when encoded in a neural network, fundamental physical properties of a BRDF, such as Helmholtz reciprocity and energy passivity, cannot be easily guaranteed. This causes undesirable effects. For example, as illustrated in \cref{fig:nbrdf-bad,fig:helmholtz}, not adhering to Helmholtz reciprocity creates tangential discontinuities around the specular region in the renderings with NBRDF models.

\section{Method}
\label{sec:method}
In this section, we introduce the formulation of our \emph{physically based neural BRDF~(PBNBRDF)}. We aim to utilize an MLP to model a BRDF that adheres to the three fundamental physical properties outlined in \cref{sec:background:brdf}. Positivity can be achieved using the rectified linear unit~(ReLU) activation function~\cite{brownlee2019gentle}. Additionally, we introduce a reparametrization layer in \cref{sec:method:helmholtz} to guarantee Helmholtz reciprocity and the chromaticity enforcement in \cref{sec:method:chromaticity} to enhance perceptual accuracy. 

However, enforcing the energy passivity constraint \cref{eq:energy} requires evaluating integrals on $f_r^{\vec{\xi}}$, which is challenging, inexact, and inefficient for neural networks~\cite[\eg,][]{davis2007methods}. To address this, we draw inspiration from AutoInt~\cite{lindell2021autoint}, which proposes fitting the derivative of the neural network to the signal of interest. Consequently, the neural network represents the anti-derivative, whose evaluation is the integral by the fundamental theorem of calculus.

More specifically, our neural network $g^{\vec{\xi}}(\vec{\omega}_i=(\theta_i, \varphi_i), \vec{\omega}_o=(\theta_o, \varphi_o))$, parameterized by $\vec{\xi}$, is trained such that the its scaled (partial) derivative fits the BRDF values:
\begin{equation}
    \label{eq:formulation}
    f_r^{\vec{\xi}}(\vec{\omega}_i, \vec{\omega}_o) := \frac{1}{\cos \theta_o \sin \theta_o} \frac{\partial^2 g^{\vec{\xi}}(\vec{\omega}_i, \vec{\omega}_o)}{\partial \theta_o \partial \varphi_o} \approx f_r(\vec{\omega}_i, \vec{\omega}_o).
\end{equation}
We denote the scaled derivative as $f_r^{\vec{\xi}}$ to simplify notation and to emphasize that the training aims to make $f_r^{\vec{\xi}}$ approximate $f_r$. We will explain in \cref{sec:method:energy} how this formulation facilitates the enforcement of the energy passivity constraint. Note that for neural networks, computing $f_r^{\vec{\xi}}$ (differentiation) can be accomplished using automatic differentiation techniques~\cite{baydin2018automatic}, which is a much easier, accurate, and faster operation than integration.

\subsection{Energy Passivity Loss via Efficient Analytical Integration}
\label{sec:method:energy}
The violation of energy passivity for a BRDF can lead to noticeable artifacts (see \cref{fig:energy_renderings}). From \cref{eq:energy}, we observe that if $\int_\Omega f_r \cos{\theta_o}\intd \vec{\omega}_o \leq 1$, the energy is absorbed, which is a valid scenario. Conversely, if $\int_\Omega f_r \cos{\theta_o}\intd \vec{\omega}_o > 1$, the energy is created. Therefore, our goal is to minimize this integral until it falls below 1. Our anti-derivative formulation allows us to find a closed-form solution for the integral as
\begin{align}
    & \int_{\Omega}  f_r^{\vec{\xi}} \cos\theta_o \intd \vec{\omega}_o \label{eq:EPL-1} \\
    =& \int_0^{2\pi} \int_0^{\frac{\pi}{2}} \frac{1}{\cos \theta_o \sin \theta_o} \frac{\partial^2 g^{\vec{\xi}}}{\partial \theta_o \partial \varphi_o} \cos\theta_o \sin \theta_o \intd {\theta}_o \intd {\varphi}_o \label{eq:EPL-2} \\
    =&  \int_0^{2\pi} \int_0^{\frac{\pi}{2}} \frac{\partial^2 g^{\vec{\xi}}}{\partial \theta_o \partial \varphi_o} \intd {\theta}_o \intd {\varphi}_o \label{eq:EPL-3} \\
    =& g^{\vec{\xi}} \bigg(\vec{\omega}_i, \vec{\omega}_o=\Big(\frac{\pi}{2}, 2\pi \Big) \bigg) - g^{\vec{\xi}}\bigg(\vec{\omega}_i, \vec{\omega}_o=(0, 0)\bigg), \label{eq:EPL-4}
\end{align}
where the transition from \eqref{eq:EPL-1} to \eqref{eq:EPL-2} is due to \cref{eq:formulation} and the fact that $\intd \vec{\omega}_o = \sin \theta_o \intd \theta_o$, and the transition from \eqref{eq:EPL-3} to \eqref{eq:EPL-4} is by the fundamental theorem of calculus. By doing this, we avoid the expensive and inexact calculation or approximation of the integral in \cref{eq:energy}. 

With this, we define the \emph{energy passivity loss (EPL)} to be
\small
\begin{equation}
\label{eq:EPL}
\mathcal{L}_\text{EPL}\left(\vec{\xi}\right) := \mathbb{E}_{\vec{\omega}_i} \left[\max \bigg(1, g^{\vec{\xi}}\Big(\vec{\omega}_i, \Big(\frac{\pi}{2}, 2\pi \Big) \Big) - g^{\vec{\xi}}\Big(\vec{\omega}_i, (0, 0) \Big) \bigg) \right],
\end{equation}
\normalsize
which we add to the original NBRDF loss to penalize the creation of energy.

\subsection{Helmholtz Reciprocity via Reparametrization} 
\label{sec:method:helmholtz}
As illustrated in \cref{fig:nbrdf-bad}, violating Helmholtz reciprocity creates undesired artifacts in NBRDF renderings. To ensure Helmholtz reciprocity, it is necessary for $f_r^{\vec{\xi}}$ to establish $\pi$-periodicity in $\varphi_{\vec{d}}$ (see \cref{sec:Rusinkiewicz Parametrization}). Therefore, in the first layer of the neural network $g^{\vec{\xi}}(\vec{\omega}_i, \vec{\omega}_o)$, we transform the inputs $\vec{\omega}_i, \vec{\omega}_o$ to the Rusinkiewicz parametrization in spherical coordinates $\theta_{\vec{h}}, \varphi_{\vec{h}}, \theta_{\vec{d}}, \varphi_{\vec{d}}$ using \cref{eq:rusinkiewicz}, and then map them into $\vec{h}$ and $\vec{d}'$ via  
\begin{align}
    \vec{h} &= (\sin \theta_{\vec{h}} \cos \varphi_{\vec{h}}, \sin \theta_{\vec{h}} \sin \varphi_{\vec{h}}, \cos \theta_{\vec{h}}); \\
    \vec{d}' &= (\sin \theta_{\vec{d}} \cos 2 \varphi_{\vec{d}}, \sin \theta_{\vec{d}} \sin 2 \varphi_{\vec{d}}, \cos \theta_{\vec{d}}). \label{eq:pi-periodicity}
\end{align}
Note the similarity of the above transformations to the standard conversion from spherical coordinates to Euclidean coordinates, except that we double $\varphi_{\vec{d}}$. The following layers of $g^{\vec{\xi}}$ operate solely on $\vec{h}$ and $\vec{d}'$ without using $\vec{\omega}_i, \vec{\omega}_o$ or $\theta_{\vec{h}}, \varphi_{\vec{h}}, \theta_{\vec{d}}, \varphi_{\vec{d}}$.

Under these transformations, it is evident that $g^{\vec{\xi}}$, and thus its scaled derivative $f_r^{\vec{\xi}}$, is $\pi$-periodic in $\varphi_{\vec{d}}$ because the only instances where $\varphi_{\vec{d}}$ appears are $\cos 2 \varphi_{\vec{d}}$ and $\sin 2 \varphi_{\vec{d}}$ in \cref{eq:pi-periodicity}, which are $\pi$-periodic in $\varphi_{\vec{d}}$. While this adjustment loses the original meaning of the difference vector from Rusinkiewicz parametrization inside the network (hence $\vec{d}'$ is used instead of $\vec{d}$), it inherently guarantees compliance with Helmholtz reciprocity at the network architecture level.
\begin{figure*}[!ht]
    \centering
    \includegraphics[width=0.74\linewidth]{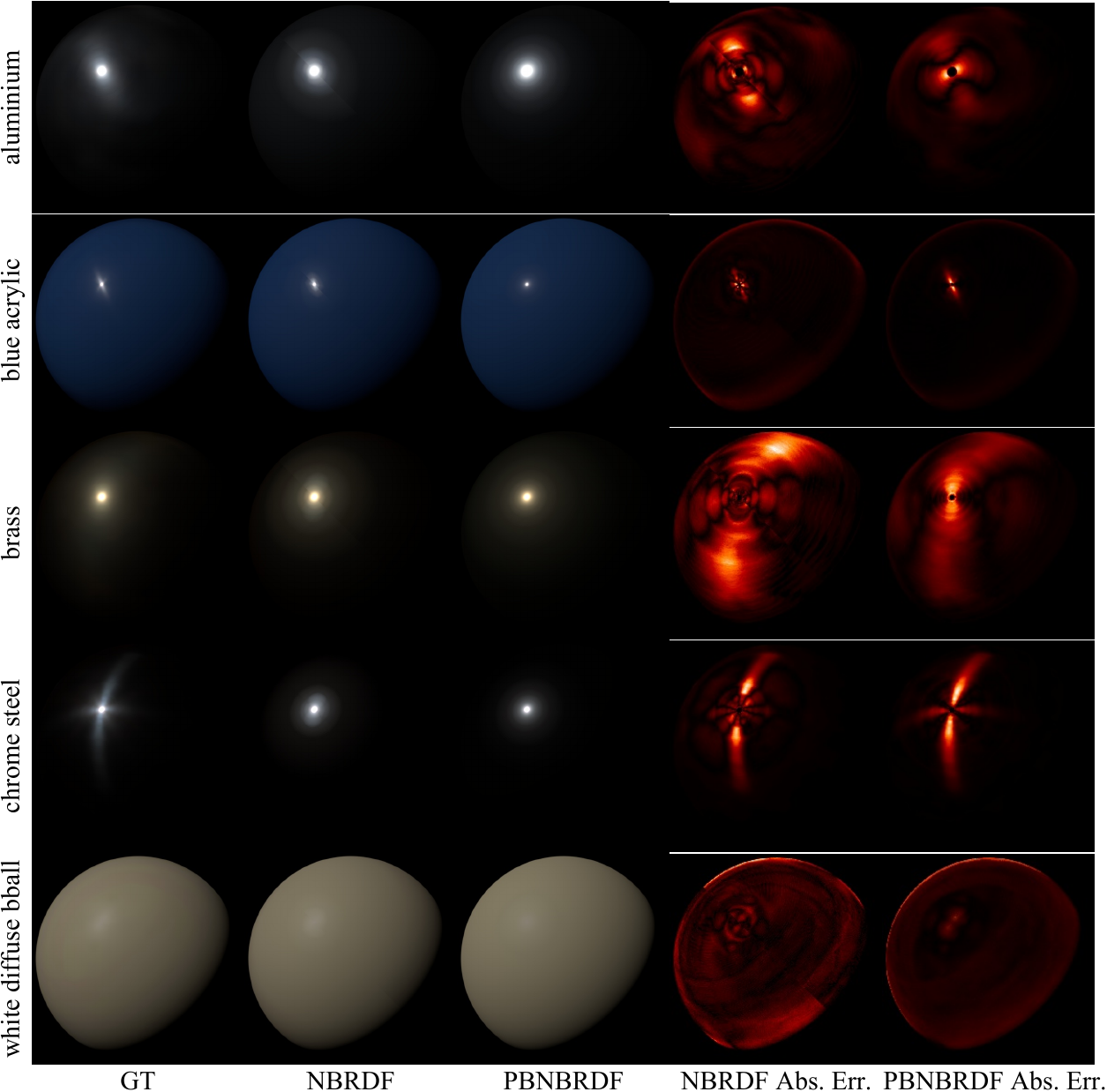}
    \caption{Renderings of NBRDF~\cite{sztrajman2021nbrdf} and our PBNBRDF fitted to MERL BRDFs~\cite{Matusik2003datadriven}, along with the absolute error plots. Violating Helmholtz reciprocity creates tangential discontinuities in NBRDF renderings, whereas our PBNBRDF produces more natural and high-quality renderings without noticeable discontinuities.}
    \label{fig:helmholtz}
\end{figure*}

\subsection{Chromaticity Enforcement via Norm Supervision}
\label{sec:method:chromaticity}
Since the BRDF values can have a very high dynamic range, reflectance value fitting is sensitive to the distribution of errors. For low values, the proportion of individual color channels can be heavily affected by small fitting errors, leading to inaccurate ``hue'' in the rendered images, resulting in unrealistic colors and a lack of visual fidelity. NBRDFs~\cite{sztrajman2021nbrdf} partly addressed these issues by using a logarithmic loss \cref{eq:nbrdf-loss} between BRDF values. While this loss ensures a more balanced consideration of low- and high-value samples during training, it is computed per color channel. Supervising the red, green, and blue channels separately proves to be unstable; for example, if the red value is slightly overestimated while green and blue are slightly underestimated, treating each color individually will result in a disproportionate shift to red at very low BRDF values, leading to a loss of visual accuracy (see \cref{fig:chromaticity_vis}).

To mitigate this inaccuracy, we introduce \emph{chromaticity enforcement~(CE)}, which enforces consistency between the individual channels indirectly by matching the norm of the predicted BRDF value with the ground truth. Specifically, we formulate the chromaticity enforcement term as a logarithmic loss on the squared norm of the RGB vector:
\small
\begin{equation}
    \mathcal{L}_\text{CE}(\vec{\xi}) := \left| \log \bigg(1 + \Big \lVert f_r \cos\theta_i  \Big \rVert ^2 \bigg) - \log \left(1 + \left \lVert f_r^{\vec{\xi}} \cos\theta_i \right \rVert ^2 \right) \right|,
    \label{eq:chroma-loss}
\end{equation}
\normalsize
where $\left \lVert \cdot \right \rVert ^2$ is the sum of squared RGB components. This encourages the error to be more consistently spread across the three channels, ensuring that the colors are perceptually closer to the ground truth. Additionally, supervising the norm of the RGB vector enhances adherence to the physical property by enforcing the total amount of reflected energy (\ie the perceived brightness) to match.

Combining the chromaticity enforcement with the energy passivity loss from \cref{sec:method:energy} and the original NBRDF training loss \cref{eq:nbrdf-loss}, the full training loss of PBNBRDF is defined as
\begin{equation}
    \label{eq:full_loss}
    \mathcal{L}_\text{PBNBRDF}=\mathcal{L}_\text{NBRDF}+ \lambda_\text{EPL} \mathcal{L}_\text{EPL} + \lambda_\text{CE} \mathcal{L}_\text{CE},
\end{equation} where $\lambda_\text{EPL}$ and $\lambda_\text{CE}$ are hyperparmeters. In practice, we set $\lambda_\text{EPL}=0.1$ and $\lambda_\text{CE}=1$.

\section{Experiments}
\label{sec:results}

\subsection{Dataset}
\label{sec:results:dataset}
\paragraph{MERL} The MERL~(Mitsubishi Electric Research Laboratories) BRDF dataset~\cite{Matusik2003datadriven} contains 100 different isotropic materials stored as densely measured BRDF values.
Both linear and non-linear dimensionality reductions are applied to discover a lower-dimensional representation that characterizes their measurements.

\paragraph{RGL} The RGL~(Realistic Graphics) dataset~\cite{rgl2018} provides a set of compact spectral BRDFs of surfaces exhibiting arbitrary roughness, including anisotropy. The materials are efficiently acquired under their adaptive parameterization using a modified goniophotometer to obtain the necessary retro-reflection measurements.
\begin{figure}[ht]
    \centering
    \includegraphics[width=0.9\linewidth]{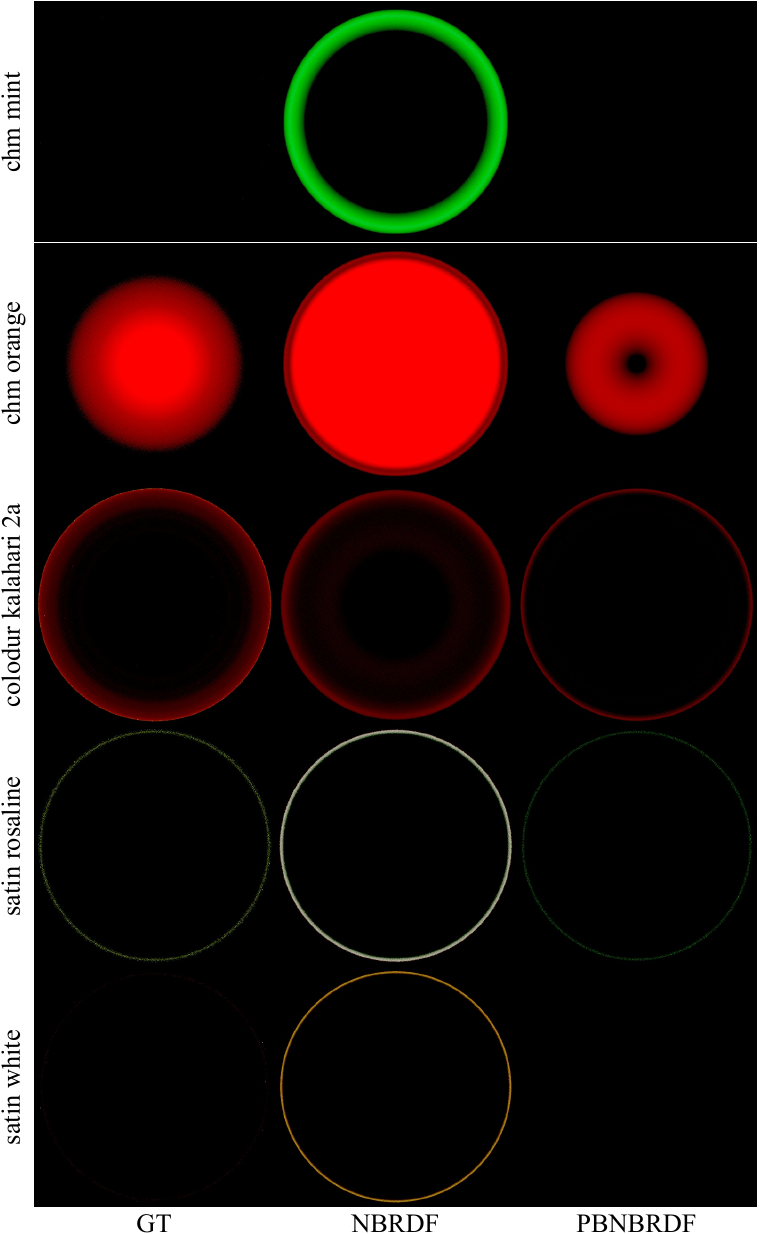}
    \caption{Renderings of NBRDF~\cite{sztrajman2021nbrdf} and our PBNBRDF fitted to the RGL dataset~\cite{rgl2018} under the white furnace test~\cite{heitz2014understanding}. The non-passivity effect is exacerbated by NBRDFs, but is significantly reduced when using our PBNBRDF\@.}
    \label{fig:energy_renderings}
\end{figure}

\subsection{Metrics} 
\paragraph{Helmholtz reciprocity metrics} To assess compliance with Helmholtz reciprocity for a BRDF $f_r$, we propose two metrics: \emph{Helmholtz reciprocity index~(HRI)} and \emph{Helmholtz continuity index~(HCI)} (to simplify notation, we assume here the inputs to $f_r$ are spherical coordinates of the half and difference vectors):
\begin{align}
    \mathcal{L}_\text{HRI}\left(f_r\right) &:= \mathbb{E}_{\raisebox{-6mm}{$\mathclap{\substack{\theta_{\vec{h}}, \theta_{\vec{d}} \sim \mathcal{U}[0, \frac{\pi}{2}] \\ \varphi_{\vec{h}}, \varphi_{\vec{d}} \sim \mathcal{U}[0, \pi]}}$}}\mkern-1mu\left[ \big(f_r(\theta_{\vec{h}}, \theta_{\vec{d}}, \varphi_{\vec{d}})
    - f_r(\theta_{\vec{h}}, \theta_{\vec{d}}, \varphi_{\vec{d}} + \pi) \big)^2 \right]; \\
    \mathcal{L}_\text{HCI}\left(f_r\right) &:= \mathbb{E}_{\raisebox{-4mm}{$\mathclap{\scriptstyle\substack{\theta_{\vec{h}}, \theta_{\vec{d}} \sim \mathcal{U}[0, \frac{\pi}{2}]}}$}}\mkern-1mu \Big[ \big| f_r(\theta_{\vec{h}}, \theta_{\vec{d}}, 0) - f_r(\theta_{\vec{h}}, \theta_{\vec{d}}, \pi) \big| \Big],
\end{align}
where $\mathcal{U}[a, b]$ is the uniform distribution on the interval $[a, b]$. HRI quantifies the degree of enforcement of Helmholtz across the entire domain of $\varphi_{\vec{d}} \in [0, 2\pi]$. However, the domain boundaries are particularly significant since any disparity between the BRDF values at $(\theta_{\vec{h}}, \varphi_{\vec{h}}, \theta_{\vec{d}}, \varphi_{\vec{d}}=0)$ and $(\theta_{\vec{h}}, \varphi_{\vec{h}}, \theta_{\vec{d}}, \varphi_{\vec{d}}=\pi)$ for given $\theta_{\vec{h}}, \varphi_{\vec{h}}, \theta_{\vec{d}}$ values might manifest as pronounced discontinuities in the rendered images, even if the difference is minimal. To address this, we introduce HCI as an additional metric to quantify such discontinuities in rendering. Both metrics aim for lower values to indicate better performance.
\begin{figure}[ht]
    \centering
    \includegraphics[width=0.88\linewidth]{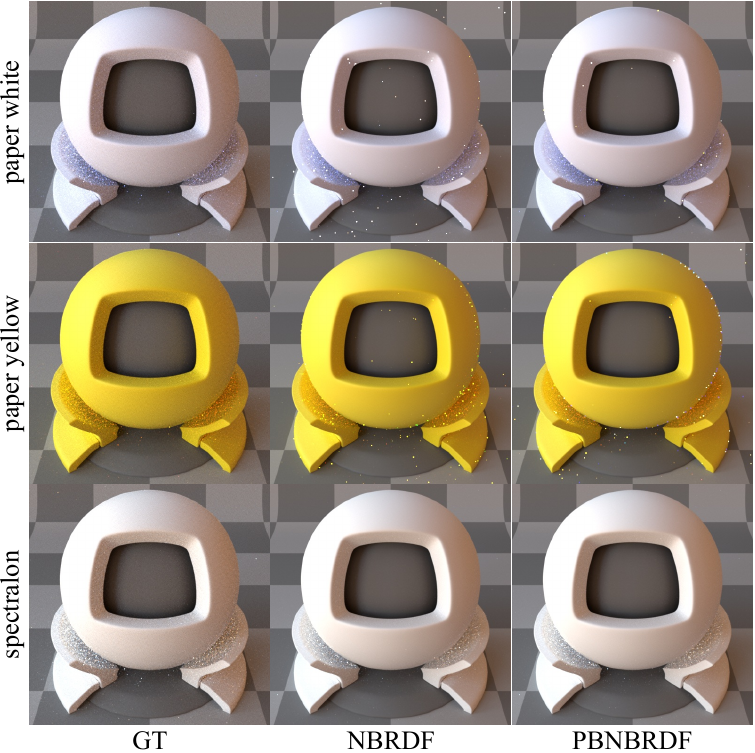}
    \caption{Renderings of NBRDF and our PBNBRDF fitted to the RGL dataset~\cite{rgl2018}. NBRDF produces noticeable artifacts (fireflies) due to energy creation, whereas our PBNBRDF provides higher-quality renderings.}
    \label{fig:more_energy_renderings}
\end{figure}

\paragraph{White furnace test for energy passivity} To visualize the violation of energy passivity in certain materials, we use the white furnace test~\cite{heitz2014understanding}, which involves illuminating a sphere with a constant white environment map of intensity 1. This setup ensures that every point on the sphere is illuminated from all directions, allowing the rendered image to visualize the integral of the BRDF slice. Every point on the sphere should evaluate to 1 if there is no energy loss or gain. By displaying such renderings after subtracting 1 from all pixel values, we can visualize only the excess light (created energy). Ideally, the visualization should be a black image, indicating no energy creation.

We also propose two numerical metrics to quantitatively evaluate energy passivity: 1. \emph{energy creation index~(ECI)}, which calculates the mean pixel value across the renderings from the white furnace test; and 2. \emph{energy passivity index~(EPI)}, which computes the proportion of reflected energy according to \cref{eq:energy}:
\begin{equation}
     \mathcal{L}_\text{EPI}\left(f_r\right) := \max \left(1, \int_\Omega f_r(\vec{\omega}_i, \vec{\omega}_o)\cos{\theta_o}\intd \vec{\omega}_o \right).
\end{equation}
Both ECI and EPI aim for lower values to indicate better energy passivity. 

\paragraph{Image metrics} We use seven metrics to assess the quality of rendered images and their similarity to the ground truths: mean absolute error~(MAE), mean squared error~(MSE), peak signal-to-noise ratio~(PSNR), structural similarity index measure~(SSIM), learned perceptual image patch similarity~(LPIPS)~\cite{zhang2018perceptual}, $\Delta E^*$~\cite{backhaus2011color}, and FLIP~\cite{andersson2020flip}. PSNR measures the quality of image reconstruction, while LPIPS, SSIM, $\Delta E^*$, and FLIP measure the perceptual difference between two images. PSNR and SSIM aim for higher values whereas the rest aim for lower values.

\subsection{Helmholtz Reciprocity}
\label{sec:results:helmholtz}
We compare NBRDF and PBNBRDF on the MERL BRDF dataset~\cite{Matusik2003datadriven} in terms of the HRI and HCI metrics for adherence to Helmholtz reciprocity, as shown in \cref{tab:small-table}, and present qualitative results in \cref{fig:helmholtz}. The negligible HRI and HCI values for PBNBRDF in \cref{tab:small-table} indicate its strong adherence to Helmholtz reciprocity. From \cref{fig:helmholtz}, we can observe that violating Helmholtz reciprocity creates tangential discontinuities in NBRDF renderings. In contrast, our PBNBRDF produces more natural and high-quality renderings without noticeable discontinuities.
\begin{table}
    \centering
    \begin{tabular}{lcccc}
        \toprule
         \multicolumn{1}{c}{Model} & HRI & HCI & ECI & EPI \\
         \midrule
        GT MERL & 0.00 & 0.00 & - & - \\
        GT RGL & - & - & 2.70 & 311 \\
        NBRDF & 14700 & 52.8 & 7.82 & 1.69 \\
        PBNBRDF & \textbf{43.3}  & \textbf{2.12} & \textbf{1.03} & \textbf{0.00} \\
        \bottomrule
    \end{tabular}
    \caption{Comparison of adherence to Helmholtz reciprocity (HRI and HCI) and energy passivity (ECI and EPI). For HRI and HCI, the models are trained on the MERL BRDF dataset~\cite{Matusik2003datadriven}, and for ECI and EPI, the models are trained on the RGL dataset~\cite{rgl2018}. All metrics are multiplied by $10^3$.
    Lower values indicate better performance. PBNBRDF outperforms NBRDF across all metrics, indicating its strong adherence to Helmholtz reciprocity and energy passivity.}
    \label{tab:small-table}
\end{table}
\begin{figure*}[!ht]
    \centering
    \includegraphics[width=0.75\linewidth]{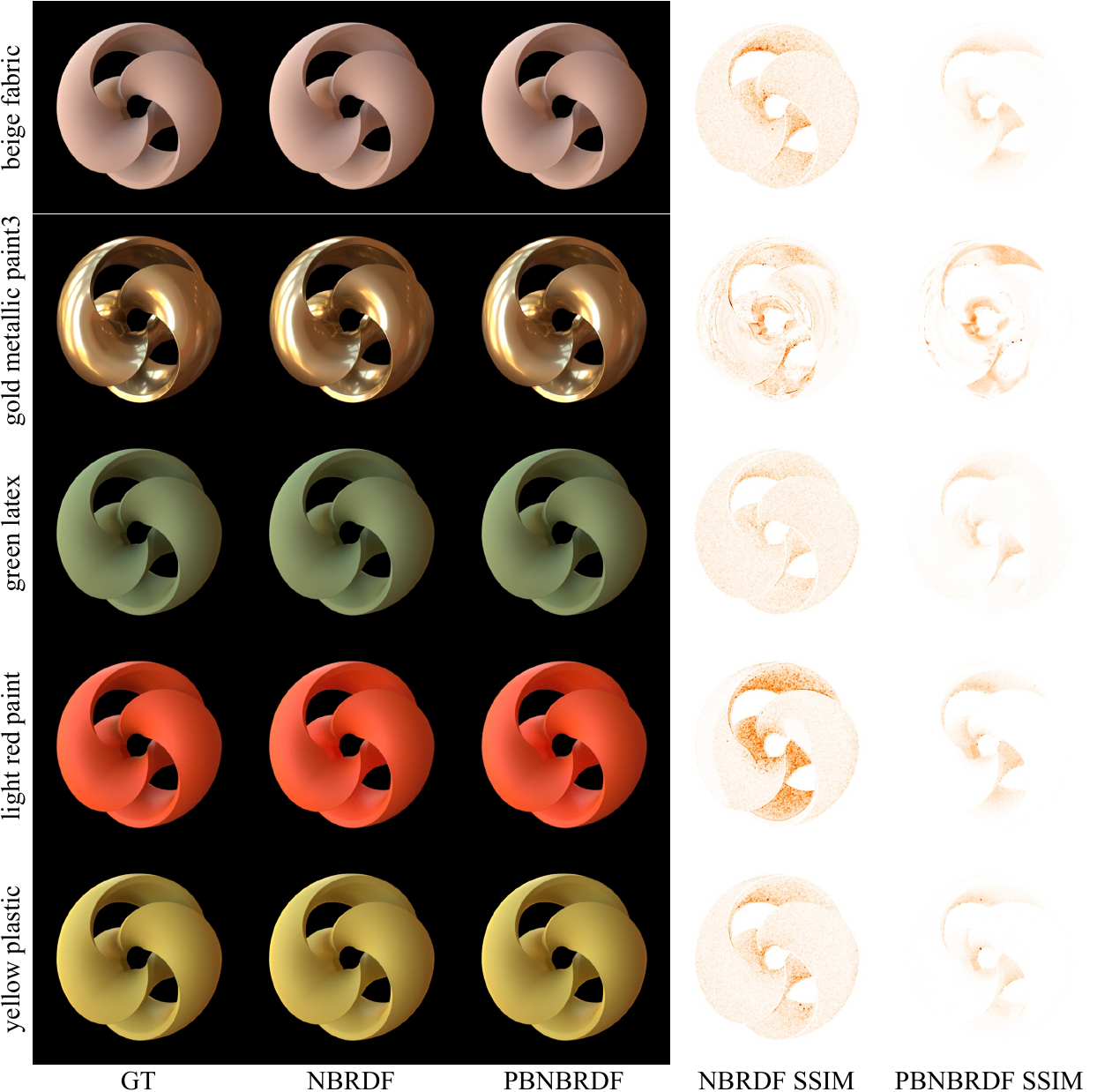}
    \caption{Renderings of NBRDF~\cite{sztrajman2021nbrdf} and our PBNBRDF fitted to RGL BRDFs~\cite{rgl2018}, along with SSIM plots. PBNBRDF produces more realistic and natural renderings that are perceptually closer to the ground truth, as indicated by higher SSIM values (shallower color).}
    \label{fig:chromaticity_vis}
\end{figure*}
\begin{table*}[!ht]


    \begin{tabular}{lccccccc}
    \toprule
        Model & MAE ($\downarrow$) & MSE ($\downarrow$) & PSNR ($\uparrow$) & SSIM ($\uparrow$) & LPIPS ($\downarrow$) & $\Delta E^*$ ($\downarrow$) & FLIP ($\downarrow$) \\
         \midrule
        NBRDF~\cite{sztrajman2021nbrdf} & 0.805 & 2.45 & 32.9 & 0.988 & 9.11 & 34.6 & 58.9 \\
        Only Helmholtz reciprocity \ & 0.620 & 1.31 & 37.7 & 0.986 & 5.43 & 97.6 & 32.0 \\
        Only energy passivity & 0.710 & 2.39 & 33.4 & 0.988 & 9.11 & 40.2 & 60.3 \\
        Only chromaticity enforcement & 0.586 & 1.31 & 36.8 & 0.992 & 6.76 & 3.25 & 58.1 \\
        PBNBRDF & 0.625 & 1.29 & 37.3 & 0.990 & 6.43 & 12.4 & 42.6 \\
         \bottomrule
        \label{tab:big-table}
    \end{tabular}
    \caption{Comparison of rendered images using NBRDF~\cite{sztrajman2021nbrdf} and different variations of PBNBRDF models on the MERL BRDF dataset~\cite{Matusik2003datadriven}. LPIPS and $\Delta E^*$ are multiplied by $10^3$. Each of the three technique for Helmholtz reciprocity, energy passivity, and chromaticity enforcement improves the quality of the rendered images. When combined, PBNBRDF consistently achieves superior performance over NBRDF across all metrics, indicating its effectiveness.}
\end{table*}


\subsection{Energy Passivity}
\label{sec:results:energy}
We compare NBRDF and PBNBRDF on materials from the RGL dataset~\cite{rgl2018}. This dataset contains BRDFs with low dissipation of energy, such as \emph{spectralon}.

In \cref{fig:energy_renderings}, we present renderings under the white furnace test~\cite{heitz2014understanding}. From the figure, we observe that the vanilla NBRDF~\cite{sztrajman2021nbrdf} creates (\emph{chm mint} and \emph{satin white}) or exacerbates (\emph{chm orange}, \emph{colodur kalahari 2a}, and \emph{satin rosaline}) the effect of energy creation. In contrast, our PBNBRDF consistently maintains energy passivity or reduces energy creation. In \cref{fig:more_energy_renderings}, we display renderings under normal lighting. We see that when fitted using NBRDF, the unconstrained regression leads to artifacts (fireflies) due to the creation of energy. On the other hand, the unsupervised energy loss EPL \cref{eq:EPL} in our PBNBRDF reduces this effect by penalizing energy creation.

These results are echoed by quantitative comparisons in \cref{tab:small-table}, where we see that PBNBRDF achieves significantly lower ECI and EPI values.

\subsection{Chromaticity Enforcement}
In \cref{fig:chromaticity_vis}, we present renderings and SSIM plots for NBRDF and our PBNBRDF models fitted to RGL materials. We observe that PBNBRDF produces more realistic and natural renderings that are perceptually closer to the ground truth, characterized by higher SSIM values (shallower color).

\subsection{Ablation Study}
\label{sec:results:ablation}
In \cref{tab:big-table}, we conduct an ablation study on the MERL BRDF dataset~\cite{Matusik2003datadriven} to provide a thorough quantitative evaluation comparing the original NBRDF method~\cite{sztrajman2021nbrdf}, our models with each of the three techniques individually for Helmholtz reciprocity, energy passivity, and chromaticity enforcement, respectively, and our proposed PBNBRDF method, which combines all three techniques. The results clearly indicate that each of the three techniques improves the rendered images and that PBNBRDF exhibits consistently higher performance across all evaluated metrics, underscoring its effectiveness.

\section{Conclusion and Discussion}
\label{sec:conclusion}
In this paper, we proposed PBNBRDF, a physically plausible neural BRDF. We ensure Helmholtz reciprocity via reparametrization of the input and ensure energy passivity via efficient analytical integration. Additionally, we introduced chromaticity enforcement to improve the perceptual quality of rendered images. Experiments have been conducted to demonstrate the effectiveness of our approach.

One limitation is that 
energy passivity is ensured by incorporating an additional loss. While this approach is effective, it would be more robust and reliable to guarantee energy passivity at a more fundamental level, such as at the network level, similar to what is proposed for Helmholtz reciprocity. Furthermore, current analyses are performed only on datasets of uniform materials; extending the ideas of neural BRDFs to more complex materials, such as spatially varying BRDFs, would be an interesting direction. We seek to explore these directions in future works.

\clearpage
\bibliography{main}

\end{document}